\documentclass[3p,times]{elsarticle}
\usepackage{color}
\begin{document}

%

%
\begin{frontmatter}

\title{First-principles study of band gap tuning in Ge$_{1-x}$Pb$_x$Se }

\author{Himanshu Lohani}

\address{Institute of Physics\\ Sachivalaya Marg, Bhubaneswar 751005,
India. \footnote{himanshu@iopb.res.in}}


\begin{abstract}
Narrow band gap and its tuning are important aspects of materials for their  technological 
applications. In this context group IV-VI semiconductors are one of the interesting  candidates.
In this paper, we explore  the possibility of band gap tuning in one of the 
family member of this family GeSe by using isoelectronic Pb doping. Our study is
first principles based electronic structure calculations of Ge$_{1-x}$Pb$_x$Se.
This study reveals that the Ge-p and Se-p states are strongly hybridized in
GeSe and shows a gap in the DOS at E$_f$ in GeSe. This gap reduces systematically with simultaneous
enhancement of the states in the near E$_f$ region as a function of Pb doping.
This leads tuning of the indirect band gap in  GeSe via Pb doping.
The results of the indirect band gap decrement are consistent with the experimental findings.
We propose a mechanism where the  electrostatic
 effect of dopant Pb cation could be  responsible for these changes in the electronic structure of GeSe.
\end{abstract}

\end{frontmatter}


\section{Introduction}
Group IV-VI semiconductors are one of the  interesting  materials from the past mainly
due to their small energy band gap. Recent studies have revealed  that these narrow energy gap materials serve 
an excellent ground for photoabsorption in near infrared and infrared regions of solar spectrum in their nano 
crystalline(NC) form which is quite useful to fabricate absorption layer of the photovoltaics\cite{Wei,hickey,lim}. 
Some of the anion alloyed NCs,  like PbS$_x$Se$_{1-x}$ and PbS$_x$Te$_{1-x}$  have already been  synthesized for  
tuning the energy gap\cite{Ma,smith} in these regions of solar spectrum. These materials also show wide variation in 
the  electrical and optical properties which make them useful in  numerous applications, for example photovoltaic
applications\cite{photo}, long wavelength sensor devices\cite{radi} and infrared lasers\cite{shi,heiss}.
Furthermore,  SnSe, GeSe, GeS, and SnS compounds of this group IV-VI family have  been recognised as
highly efficient materials for thermoelectricity because of their large Seeback  coefficients, high power factor and low
thermal conductivities\cite{thermo}. Intermediate compounds of this family have also been formed
for further investigating the properties of these materials. For example, currently Gharsallah {\it et al.} have claimed 
that Ge doping can enhance the thermoelectric efficiency of SnSe. They argue increase in the band gap could be the reason 
for record high Seeback coefficient observed in the lighter doped Sn$_{1-x}$Ge$_x$Se system\cite{thermo1}.

GeSe is one of the member of this family of semiconductors. It exhibits an orthorhombic crystal structure 
where Ge and Se atoms are arranged   in two adjacent double layers. Interaction between the adjacent  layers
is weak which provides GeSe structure to a quasi two dimensional(2D) character\cite{Okazaki,gese}. 
This layered  structure leads to  various novel properties, like  
high anisotropy in optical response \cite{Ming},
drastic reduction, almost five times,  in resistivity  at high pressure(60 Kbar)\cite{Vass,Bhatia}. These properties
are  interesting for fundamental physics point of view, as well as for technological applications.
On the other hand,   PbSe  differs substantially from GeSe, though PbSe also belongs to
the same group IV-VI family of compounds. It shows  an ionic bonding and cubic NaCl
type structure\cite{Littlewood,Arthur}. The experimentally observed energy band gap of PbSe $\sim$ 0.165 eV\cite{dalven} is
also significantly smaller compared to GeSe $\sim$ 1.07 eV\cite{otto}. Furthermore, the rocksalt
structured PbSe type group IV-VI semiconductors have been recently emerged for hosting new topological state of
matter defined as "topological crystalline insulator(TCI)". Experimentally this TCI state has  been  realized in materials 
 like, Pb$_{1-x}$Sn$_x$Se\cite{tci}. The reason behind the
different nature of the group IV-VI compounds is  mainly inherent to the stereochemical activity of group IV elements as
elucidated in a comprehensive first principles study on these materials by Wagmare {\it et. al.}\cite{waghmare}

Recent experimental studies have shown that the band gap of bulk GeSe  can be tuned by isoelectronic
Pb doping\cite{thesis,thesis1}. Similar results have also been observed
in Ge$_{1-x}$Se$_2$Pb$_x$ thin films where reduction  of non direct optical gap has been observed  with Pb doping \cite{thin}. 
In our previous study, we  found that our first principles based calculated bands show a 
fair resemblance with our angle resolved photoemisson spectroscopy(ARPES) results on GeSe\cite{arpes}. 
In this paper, we extend our first principles study to explore the effects of isoelectronic Pb doping in 
the electronic structure of GeSe compound.
In band structure calculations, we find a systematic reduction of the  energy gap in GeSe with Pb doping.
This trend of indirect band gap reduction   as a function of Pb doping shows a qualitative similarity with its
 behaviour observed in the experiment. We discuss these doping induced changes in the light of electrostatic effects 
arising from the strong electropositive character of the dopant Pb atom. This study elucidates the possibility  of
the band gap tuning in GeSe compound via Pb doping.

\section{Details of the Calculation}
First-principles calculations were performed by using {\it {\textcolor{red}{planewave}}} basis set inherent in Quantum Espresso(QE)\cite{qe}.
Generalized gradient  approximation(GGA) with Perdew-Burke-Ernzerhof(PBE)\cite{Er,Wang,Chevary} parametrization 
was used to represent many electron exchange-correlation energy. It was implemented in a fully relativistic nonlinear
core corrected, ultrasoft  pseudopotential\cite{Vanderbilt,not} of Ge, Se and Pb atoms. The kinetic energy and charge density
cut-off were set to 60 Ry and 600 Ry  respectively. Mesh of 13$\times 8\times$5  Monkhorst- Pack k-points with Gaussian smearing
of the order 0.0002 Ry was used for sampling the Brillouin zone integration\cite{monk}. Experimental lattice parameters and 
atomic coordinates were used to construct a supercell of dimension a$\times 2b\times$c\cite{Okazaki}. This structure was relaxed  under
damped (Beeman) dynamics  with respect to both ionic coordinates and the lattice vector for parent and all
doped cases. All parameters were optimized under several convergence test and  change in the total
energy less than 10$^{-7}$ eV was the criteria of convergence for self consistent calculations(SCF). Wannier functions were obtained by
using standard parametrs of wannier90 code\cite{mostofi}. In this process of Wannier function construction
a dense k mesh of  10$\times 6\times$4 with convergence tolerance
of 1.00 $\times$ 10$^{-10}$ was employed. The imaginary part of these Wannier functions was less than 10$^{-3}$ to their corresponding
real part which ensured that the Wannier functions were real.

\section{Results and discussion}
\begin{figure}
\includegraphics[width=8cm,keepaspectratio]{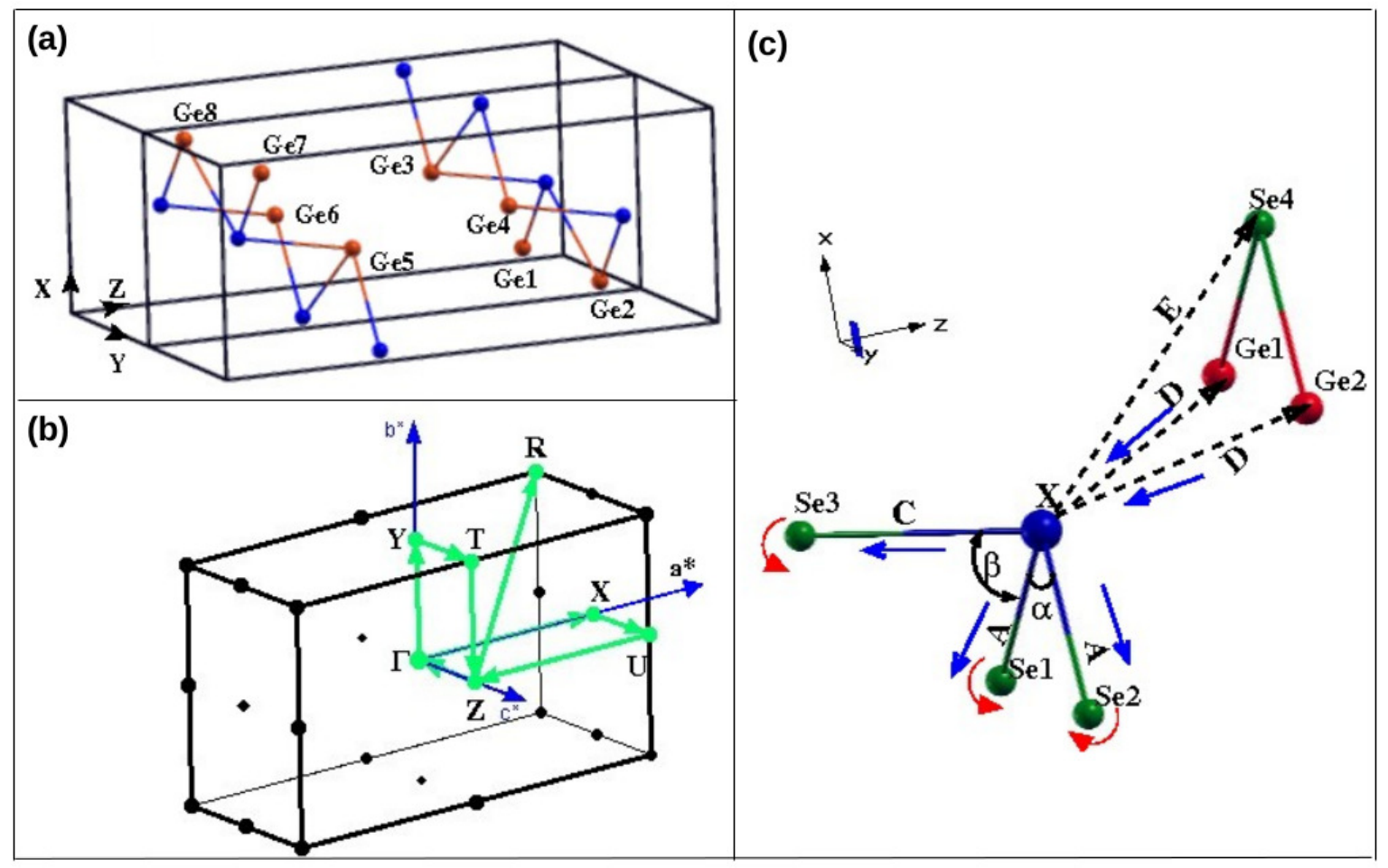}
\caption{\label{gesecrystal}(a) Two unit cell of GeSe crystal.
Red and blue dots represent the Ge and Se atoms respectively which are
arranged in two adjacent double layers. (b) Brillouin Zone of GeSe where Green colour arrows
show the k-path along which bands are calculated. (c) Coordination of cation X(Ge) in GeSe. Red and blue arrows
mark the changes in bond angles and bond lengths after substitution of Pb atom at the X site.} 
\end{figure}

Fig.\ref{gesecrystal}(a) shows two units of orthorhombic unitcell of GeSe crystal.
This crystal belongs to space group symmetry D$_{2h} ^{16}$ Pcmn(62) and its  lattice parameter
c = 10.78 \AA{} is  larger than other two lattice constants a = 4.38 \AA{} and b = 3.82 \AA{}\cite{Okazaki}.
In the  unit cell Ge/Se atoms  are arranged in two adjacent double layers where each layer consists of two double
 corrugated planes.  The  coordination environment of the Ge atom is 
illustrated  in Fig.\ref{gesecrystal}(c). It can be seen that  two first nearest neighbour (fnn) Se atoms(Se1 and Se2)
 and second nearest neighbour (snn)  Se atom (Se3) are situated at the comparable distances ($\sim$ 2.56 \AA{}) from the Ge atom.
Whereas,  third nearest neighbour (tnn) Se atom (Se4) and two fnn  Ge atoms (Ge1 and Ge2) reside on the adjacent double layer with
interatomic distances of $\sim$ 3.38 \AA{}. Bond angles $\alpha$ and $\beta$ formed by the atoms  Se1-Ge-Se2 and Se3-Ge-Se1/Se2 are
  95$^\circ$  and 90.8$^\circ$  respectively. In doped compounds  Ge atoms are substituted by Pb atoms in the supercell structure of GeSe.
In order to choose the substitutional sites for Pb doping, total  SCF energy 
was compared of all the possible  supercell structures of 0.25 doped GeSe which can be formed by substituting  two   Pb
atoms at different Ge sites. This was  minimum for the structure, where both the Pb  atoms were situated  maximally apart from each other. 
To keep this fact in mind Pb atoms were substituted in the supercell structure of GeSe (Fig.\ref{gesecrystal}(a)) at Ge1; at Ge1, and Ge5;
at  Ge1, Ge4, and Ge8 and at Ge3, Ge4, Ge7, and Ge8 for doping concentrations x = 0.125, 0.25, 0.375, and 0.5   respectively.
And cohesive energy was calculated for all these structures which informed their stabilization in the orthorhombic phase. Similarly,
structural relaxation of all the structures  infers that mainly the coordination geometry is  affected around the doped atom sites.
 The changes one of such site are marked by red and blue arrows in Fig.\ref{gesecrystal}(c). The distance of fnn and snn Se atoms 
with respect to the dopant atom (Pb)  increases $\sim$  0.2 \AA{} whereas a quite small reduction of $\sim$ 0.05 \AA{} is observed in
 the distance of tnn Se atom(Se4). The bond length of Ge1-Pb and Ge2-Pb decreases $\sim$ 0.1 \AA{} which probably  enhances the interaction between
two adjacent double layers. Similarly, the bond angles $\alpha$, and $\beta$ are reduced by $\sim$  3$^\circ$ which
brings the fnn and snn Se atoms  close to each other and thereby  distorts  each double layer locally. These 
changes in the coordination geometry around the dopant site have been compared in all the compositions  at different substitutional sites.
and found that the changes remain almost same in all the cases. 

\begin{figure}
\includegraphics[width=8cm,keepaspectratio]{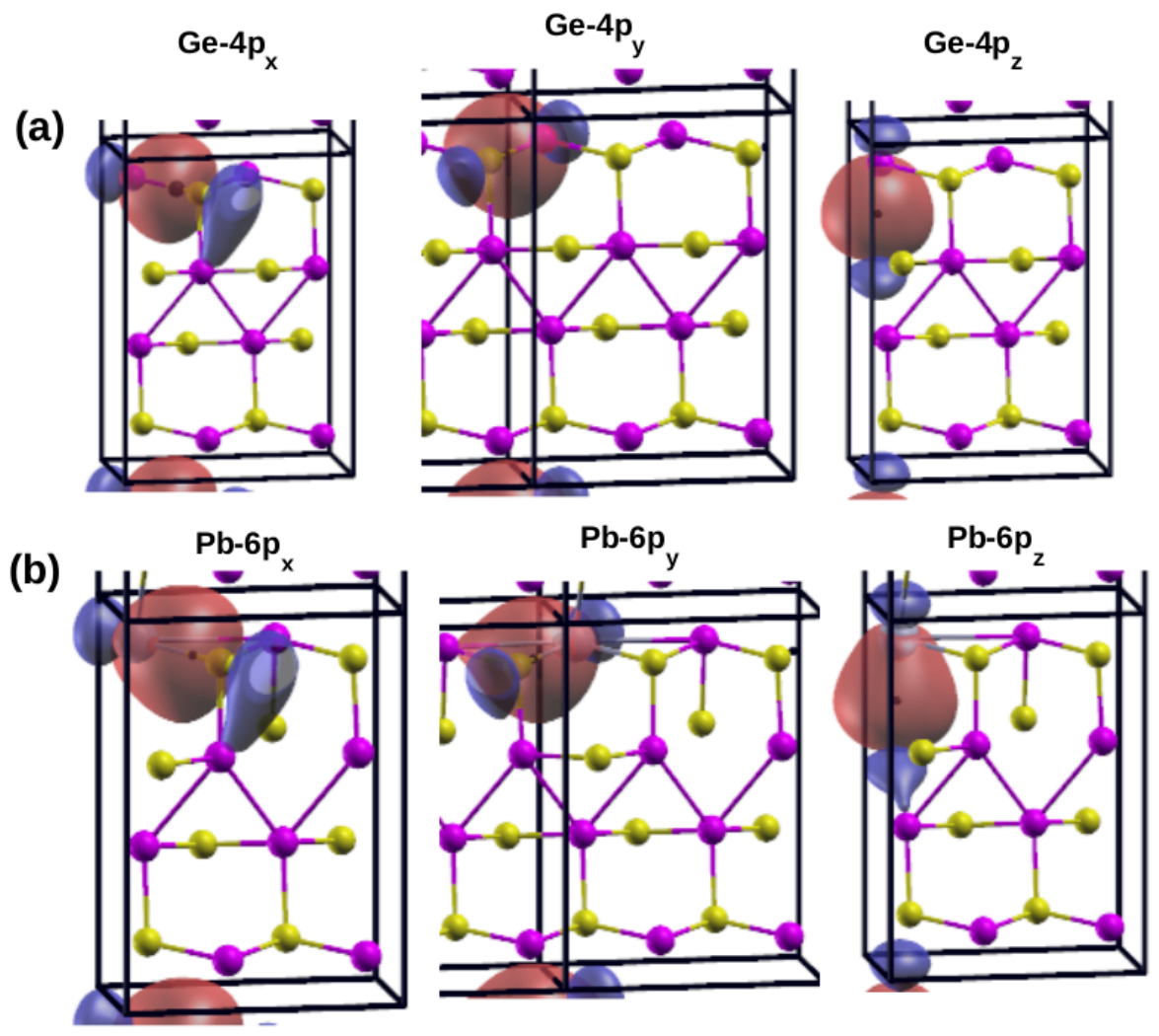}
\caption{\label{wannier} Panel (a) and (b) show  Wannier functions of Ge-4p and Pb-6p orbitals in GeSe and Ge$_{0.75}$Pb$_{0.25}$Se 
respectively, where magenta, and yellow balls correspond to Ge and Se atoms respectively, and black colour dot represents
bond center. The isosurface value is one in these Wannier functions.}
\end{figure}

Next, we present results of Wannier functions to understand the bonding character in these compounds. Panel (a) of Fig.\ref{wannier}
shows Wannier function  of Ge-4p$_x$, 4p$_y$, and 4p$_z$, where black dot represents  bond center.
These Wannier functions are elongated more towards the fnn Se atoms which is also reflected in the position
of their bond center. This provides a slightly  polar character to the covalent bonding in GeSe and
could be the result of higher electronegativity of Se atom in comparison to  Ge atom. 
Panel (b) of Fig.\ref{wannier} shows Wannier functions (Pb-6p$_x$, 6p$_y$, and 6p$_z$)  of dopant Pb atom where
an additional shift   can be seen in the position of bond center of the  Pb Wannier functions
towards the fnn Se atoms in comparison to the situation in GeSe compound. This is an indication of  further enhancement
of  polar character of the covalent bonds  at the Pb doapnt sites compared to the sites of Ge atom.

\begin{figure}
\includegraphics[width=10cm,keepaspectratio]{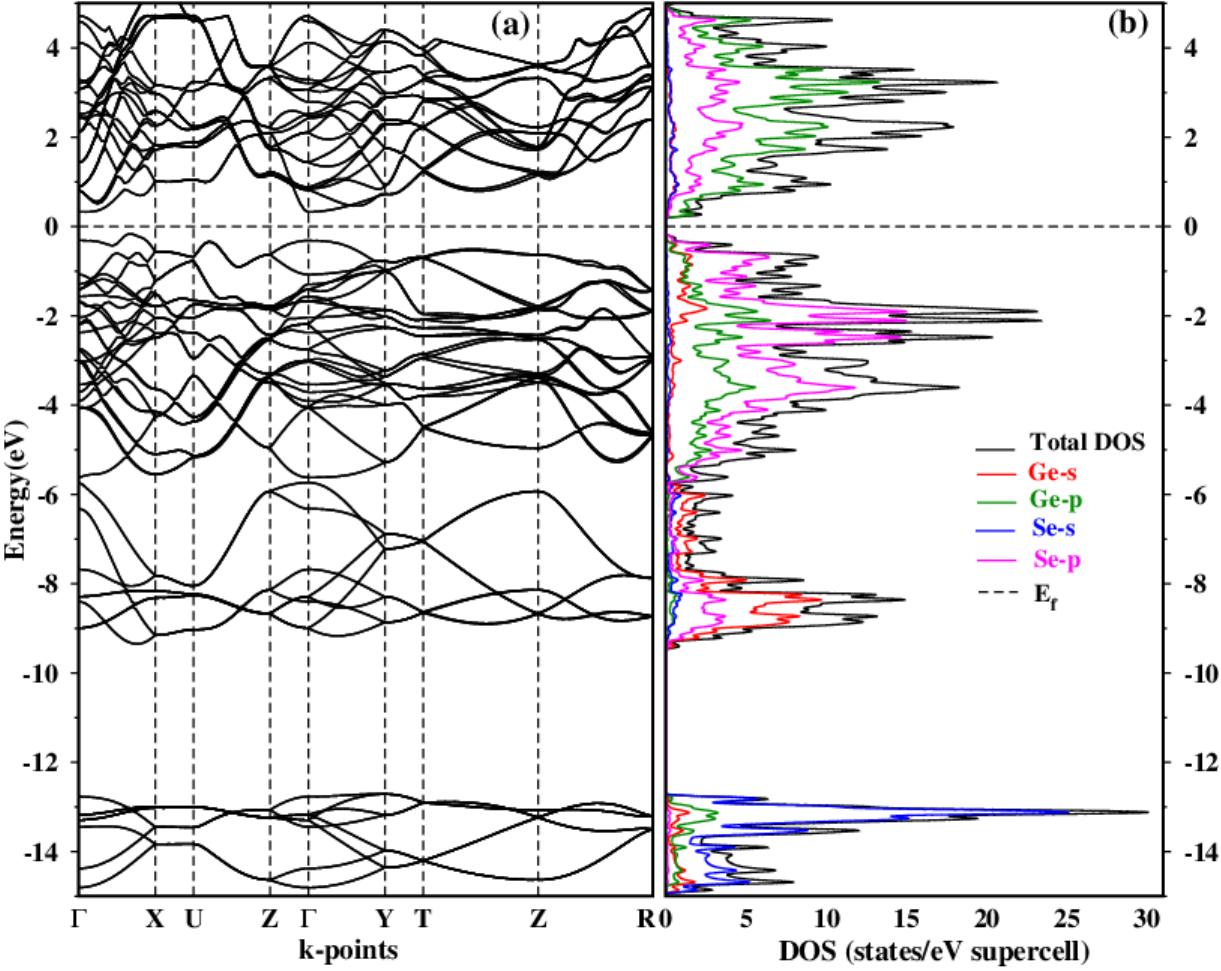}
\caption{\label{geseband} (a) Band structure plot of  GeSe  structure
(b)Total DOS of GeSe along with contribution of different orbitals of Ge and Se atoms.}
\end{figure}

Fig.\ref{geseband}(a)  shows the band structure plot of GeSe. In the VB, bands are arranged
in three separated energy regions,  -14.47 to -12.48 eV; -9.32 to -5.52 eV; and 
-5.5 eV up to the E$_f$. Similarly the CB is also gaped from the VB with direct band gap(at the $\Gamma$ point) of 0.63 eV.
However, possibility of an indirect band gap(0.49 eV) transition is also visible  from the VB maximum near the X point
 to the CB minimum at the $\Gamma$ point.
The bands are dispersive along the $\Gamma$-Z and U-X directions mainly in the BE range E$_f$ to -6.0 eV.
These directions are perpendicular (k$_z$) to the plane of the adjacent double layer(Fig.\ref{gesecrystal}(b)). This
k$_z$ dispersion of the bands suggest a moderate hybridization between the adjacent double layers in the system.
 Similarly, the degenerate bands
 directed along the U-Z direction get splitted  along the $\Gamma$-X k-path which represents the   k-direction
same as the U-Z but  lies on the adjacent double layer. In  Fig.\ref{gesedos}(b)  total density of states(DOS)
 of GeSe with different atomic contributions is plotted. Three demarcated regions in the  DOS structure are consistent
with the band picture. From partial DOS it is clear that the states are primarily composed of hybridized states between 
Se-4p and Ge-4p  in the energy range -6.0 to 4.5 eV  with an energy gap of 0.49 eV  at Fermi level(E$_f$). 
In this energy window the VB and CB parts are dominant by Se-4p and Ge-4p states and they contribute $\sim$
62.57\% and 59.89\% to the total DOS in the VB and CB regions respectively. On the other hand,  lone pair states of Se-4s and Ge-4s 
reside deep in the BE in a narrow energy interval of  $\sim$ -15 to -13 eV and -9.5 to -8.2 eV respectively.
The lone pair states of Se-4s are quite separated from the 4p hybridized states of Se and Ge,  while the Ge-4s states(7.4\%) show a little 
admixture with these hybridized states. The Ge-4s states also contribute at the top
 of VB region which  is defined as stereochemically active nature of the Ge-4s lone pair.
These states have been found to involve in secondary bonding(s-s interaction) and played
 an important role in the formation of near  E$_f$ DOS in group IV-VI compounds as reported previously\cite{Andriy}.
These results of  band and DOS calculations  on GeSe match fairly  with previous report of Albanesi {\it et. al.}\cite{gese}.

\begin{figure}
\includegraphics[width=10cm,keepaspectratio]{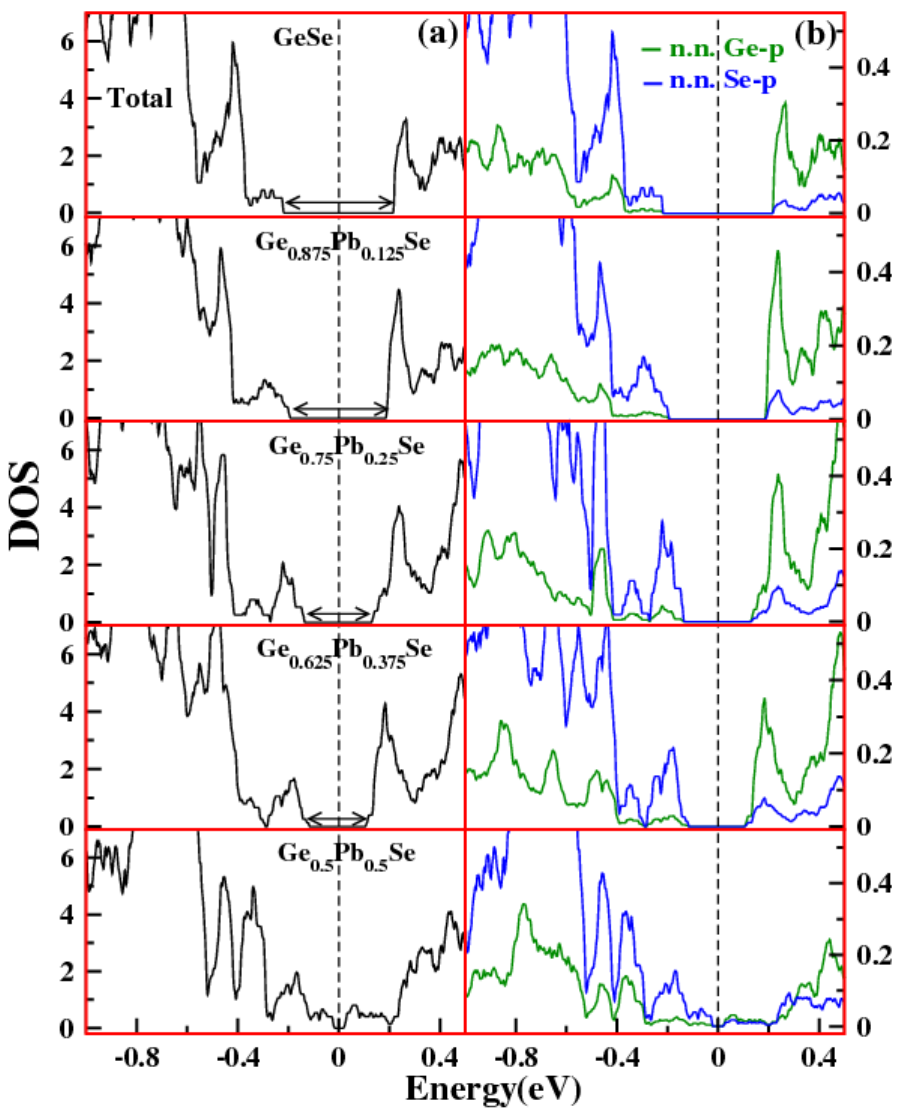}
\caption{\label{pbdos} (a) Variation in total DOS of GeSe with Pb doping(top to bottom). 
Separation between valence and conduction band edges decreases with increase of Pb doping,
which is marked by double sided arrow of black colour. (b) Changes in the contribution  of 
first nearest neighbour Ge-4p and Se-4p orbitals of the 
doped atom site with  Pb doping(top to bottom).}
\end{figure}

Next, we compared  DOS of the Pb doped GeSe compounds and found interesting changes in the near E$_f$ region of  the DOS
as depicted in Fig.\ref{pbdos}(a). Both the VB and CB edges gradually shift towards the 
 E$_f$ as a function of Pb doping.
 Apart from reduction in the energy gap, the states are also  significantly enhanced
just below and above the E$_f$ which  could be helpful in the increment of the photoabsorption in this system.
As discussed previously, with Pb doping  mainly the coordination environment affects  around the doped  sites.
So, DOS composed of the fnn atoms of doped atom probably contribute primarily to the modified DOS in GeSe due to  Pb doping.
This view is supported by Fig.\ref{pbdos}(b), where DOS of the fnn Ge-4p and Se-4p orbitals of the doped atom are plotted
in different doped compounds. As it is clear from this figure that these fnn Ge-4p and Se-4p  states in each composition highly
 resemble with the corresponding total DOS and also show similar trend  of the energy gap decrement  at the E$_f$  as a function of  doping.

\begin{figure}
\includegraphics[width=15cm,keepaspectratio]{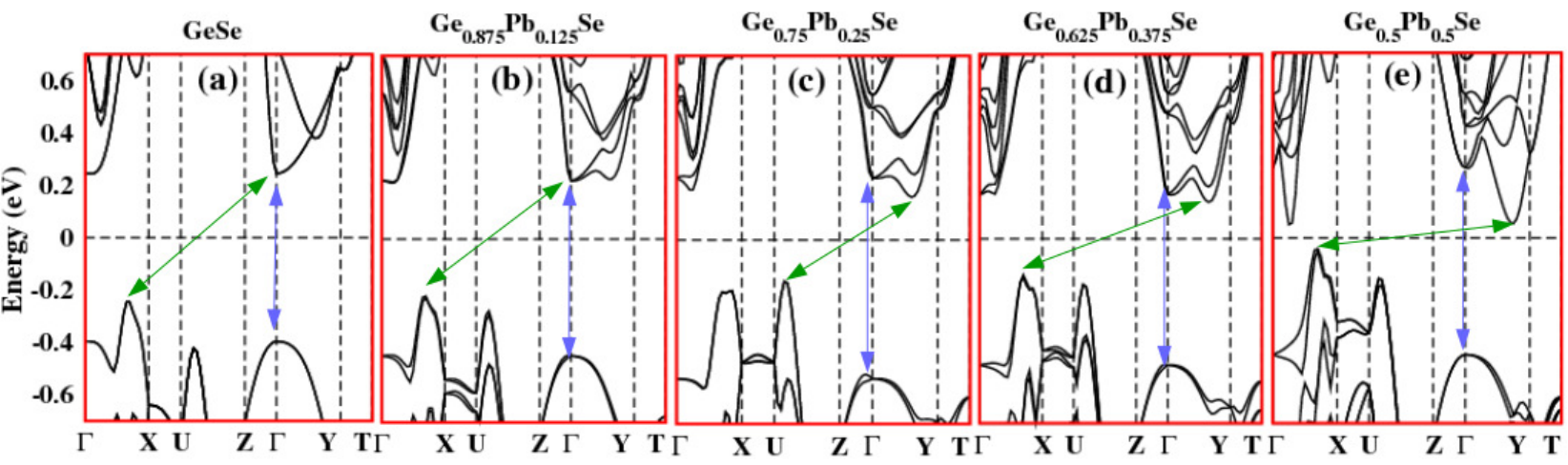}
\caption{\label{pbcom} (a) to (e)  Band structure of GeSe with different amount of Pb doping,
where blue and green arrows indicate direct and indirect band gap transitions respectively.
The direct and indirect band gap of GeSe is 0.63 eV and 0.49 eV respectively.}
\end{figure}

Band structure of different doped compounds  informs
that the bands close to Fermi level are mainly affected with doping. The
near E$_f$ band structure of various compositions are displayed in Fig.\ref{pbcom}(a)-(e),
where it can be viewed that  the lowest CB along the $\Gamma$-Y direction
gradually comes down towards the E$_f$ with the  Pb doping. While  the top most VB along the  $\Gamma$-X and U-Z directions
gradually goes up towards the E$_f$. This opposite moment of the CB and VB edges leads to a  reduction in  the indirect
band gap (marked with green arrow).
On the other hand CB edge at the $\Gamma$ point which involves in direct band gap
show slight shift towards the higher BE for 0.125 case compared to the undoped case. In the
next doping(0.25) the CB shifts slightly back to the lower BE then again comes back to the higher BE in 0.375 case.
And in the  0.5 doped case it shows a significant shift towards the higher BE. Similarly,
the VB edge at the  $\Gamma$ point shows a small drift towards the higher BE from doping doping concentration x = 0.0 to 0.25
and opposite movement {\it i. e.} higher to lower BE for 0.375 to 0.5. Thus these changes in the CB and
VB edges at the $\Gamma$ point provide  an almost constant value of the direct band (marked with blue arrow)
with Pb doping.

\begin{figure}
\includegraphics[width=12cm,keepaspectratio]{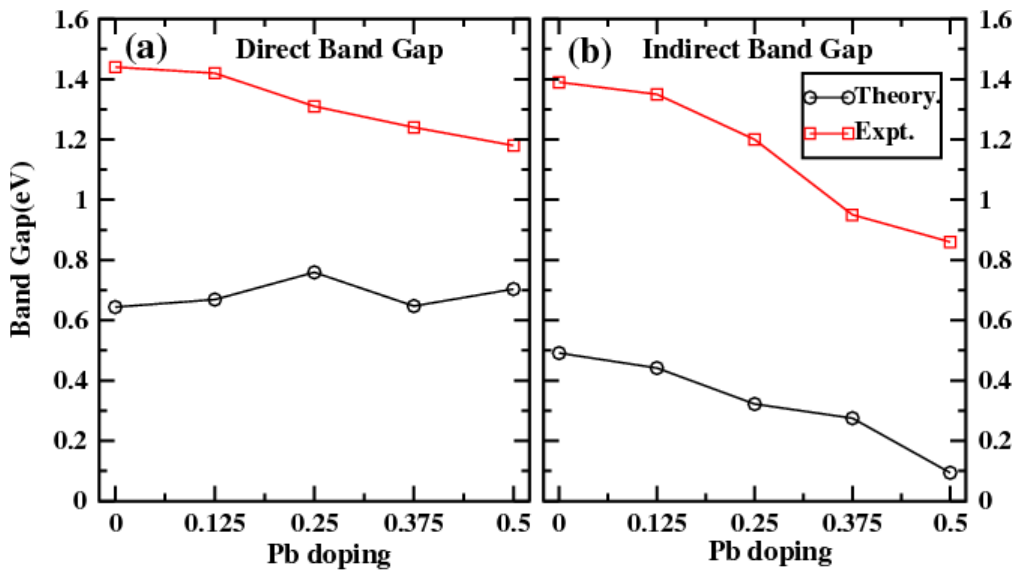}
\caption{\label{bandgap} plot (a) and (b) show changes in the direct and indirect band gap with
Pb doping, where black circle and red square denote  calculated and experimentally measured values respectively.}
\end{figure}

In Fig.\ref{bandgap}(a) and (b), we compared the calculated values of direct and indirect band gaps
of compositions Ge$_{1-x}$Pb$_x$Se (x = 0.0, 0.125, 0.25, 0.375 and 0.5) with experimental values of
 compositions Ge$_{1-x}$Pb$_x$Se (x = 0.0, 0.1, 0.2, 0.3 and 0.4) respectively. 
The trend of indirect band gap reduction found in the calculation  matches qualitatively to the experiment.
There is  an energy  difference of  $\sim$ 0.08 eV   between both the  values. On the other hand,
the calculated value of direct band gap  remains almost constant with doping, unlike the experimental
value which shows slight decrement   as a function of doping. This discrepancy could be arised
from approximation used to formulate the many body electron interaction part in the DFT calculation.
In this situation,  the present first principles study could provide a ground for further 
experimental and theoretical investigations in this Pb doped GeSe and related systems to resolve
the discrepancy of the energy band gap.

\begin{figure}
\includegraphics[width=10cm,keepaspectratio]{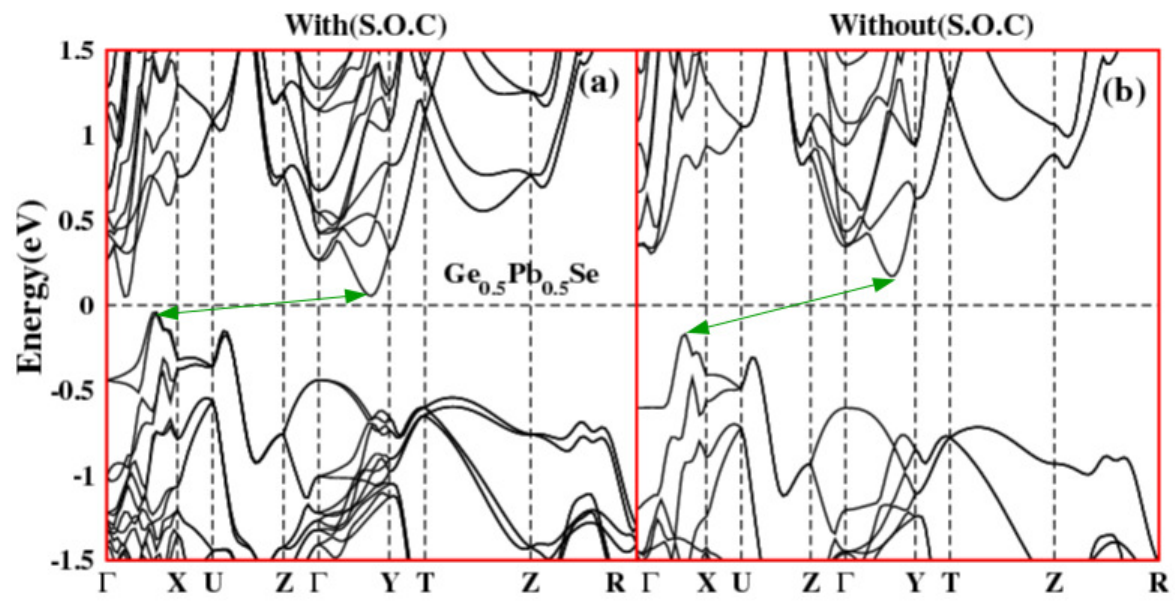}
\caption{\label{soccom}(a) and (b)  band structure plots of Ge$_{0.5}$Pb$_{0.5}$Se with and without
including the  spin orbit coupling(SOC) respectively,
where green arrow shows the indirect band gap transition.}  
\end{figure}

The effects of SOC are important in the  elements which have a high atomic number(Z), like Pb. Because 
it is proportional to the atomic number of the elements (Z$^4$). Therefore, the role of SOC effects
are  examined in the case of highest Pb concentration (0.5). 
Fig.\ref{soccom}(a) and (b) show  band structure of Ge$_{0.5}$Pb$_{0.5}$Se 
without and with inclusion of SOC effects respectively.  Comparing the SOC results
to non SOC case, a significant splitting can be viewed in the bands  due to SOC, especially
along the $\Gamma$-X and $\Gamma$-Y directions. Likewise, both the CB and VB edges move remarkably
towards the E$_f$  under the influence of SOC effects. This gives rise to lowering of the indirect band gap
in SOC case in comparison to the non SOC case. These results signify the vital role
of SOC  effects in the modification of the electronic structure of Ge$_{0.5}$Pb$_{0.5}$Se. 
The important role of SOC effects in various binary and quaternary compounds of group
IV-VI family, like Bi$_2$Se$_3$, B$_x$Sb$_{1-x}$Te$_x$Se$_{1-x}$\cite{topo}
has also been  discovered recently in the formation of  non trivial topological surface states.

\begin{figure}
\includegraphics[width=8cm,keepaspectratio]{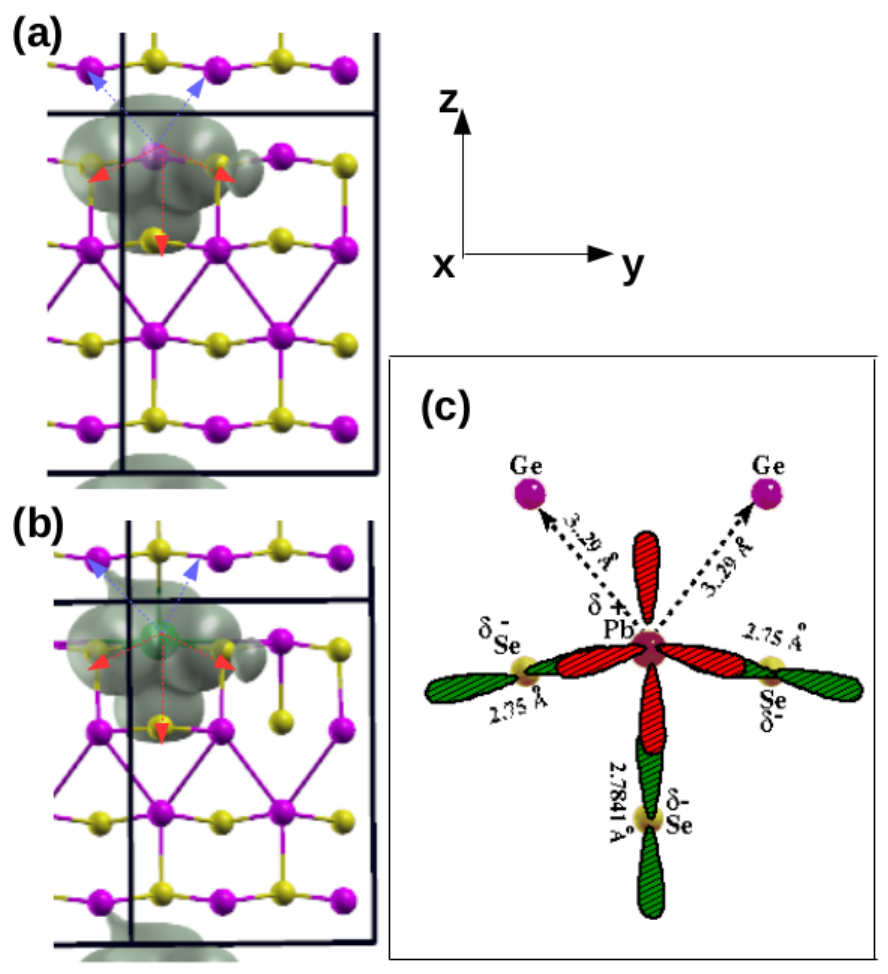}
\caption{\label{mnism} (a) and (b)  charge density of Ge and Pb atom
 in GeSe and Ge$_{0.875}$Pb$_{0.125}$Se  obtained by using their 4(s,p) and 6(s,p) Wannier functions
respectively, where magenta and yellow dots correspond to Ge and Se atoms while red and blue dotted arrow mark the first
 nearest neighbour Se and Ge atoms respectively from the  Ge/Pb atom. 
The isosurface value is five in these plots.
(c) Schematic picture of local coordination at Pb atom site in  Ge$_{0.875}$Pb$_{0.125}$Se 
compound. Green and red colour lobe represent the p orbitals of doped  Pb and its nearest neighbour Se atoms respectively.
The dotted arrows guide the position of first nearest neighbour Ge atoms form the doped Pb atom site.}
\end{figure}

In order to figure out the reason behind the gradual shifting of the VB and CB edges, we calculated the charge  
density of   Ge atom in GeSe  and doped Pb atom in  Ge$_{0.875}$Pb$_{0.125}$Se. These charge densities were
calculated by using their s and p Wannier functions and results are displayed in  Fig.\ref{mnism}(a) and (b) respectively.
In both the cases, the charge density covers the fnn Se atoms while the fnn Ge atoms are outside this coverage.
In addition, bond center of the dopant Pb-6p Wannier functions lies close  to its fnn Se atoms as clear from the Fig.\ref{wannier}(b). 
This provides a positive character ($\delta^+$)
to the Pb atom and a negative character ($\delta^-$) to the fnn Se atoms.
Thus, doped site acts like a positive charge center and due to its electrostatic interaction it 
tries to pull all the bands towards the higher BE.
The effect of this positive charge of the doped atom  should be more to its nn atoms due to
inverse square dependency of the Coulomb interaction with the distance from the source. Therefore, the fnn 
Ge-p$_{x/y}$ orbitals pull down to the higher BE under the influence of Pb dopant. However,
 energy of the fnn Ge-p$_z$ orbital should not change appreciably because   this orbital
is lesser involved in the covalent bonding compared to the fnn Ge-p$_{x/y}$ and thereby
it readjusts itself according to the new environment. It could be possible due to one of the
the fnn Ge-p$_z$ lobe( directing out of the fnn Ge-p$_{x/y}$ plain) is more distant from the nn Se atoms.
On the other hand, in the case of fnn  Se-p$_{x/y}$ orbitals the effect of charge  at the dopant site is
substantially screened by the bonding electron cloud  which is resided close to these fnn Se atoms.
Further, the electronic repulsion of these bonding electron cloud pushes the fnn Se-p$_{x/y}$ orbitals
to the lower BE. The same situation could also persists for the fnn Se-p$_{z}$ orbital but it is also
oriented out of plain, like the fnn Ge-p$_{z}$. So, there is possibility of readjustment of this orbital
to reduce the repulsion of the  bonding electronic cloud. As it is clear from the DOS plots(Fig.\ref{geseband}(b))
that the modified DOS structure in the VB and CB region (in the vicinity of E$_f$) with Pb doping is mainly 
composed of the nn Se-p and Ge-p atoms of the doped atom. This informs that the bands present close to
the E$_f$ in the VB and CB region are dominant with the fnn Se-p and Ge-p  atoms of the doped atom.
Therefore, the shifting of the VB edge towards the lower BE along the U-X and  $\Gamma$-Y directions as a function of Pb doping
is the result of electronic repulsion seen by the fnn Se-p$_{x/y}$ orbitals. Similarly,
shifting of the CB edge towards the higher BE along the $\Gamma$-Y direction could be the outcome of electrostatic attraction 
exerted on the fnn Ge-p$_{x/y}$ orbitals by  the doped atoms. 
The VB edge shows a small shifting to higher BE energy at the $\Gamma$ point for doping
x = 0.0 to 0.25. However, this VB edge moves
slightly towards lower BE for compositions x= 0.375 and 0.5. In these highly doped systems, two Pb atoms are present 
in one of the double layers (at dopant site Ge1 and Ge5 in x = 0.375; Ge1/Ge2, Ge5/Ge7 in x = 0.5), unlike the cases of 
x = 0.125 and 0.25 doping. These atomic arrangements probably modify the bonding electron cloud at the doped atom site
such that electronic repulsion overcomes the attractive electrostatic interaction of doped Pb cation ($\delta^+$)
and thereby moves the fnn Se-p$_{z}$ bands to lower BE, like the Se-p$_{x/y}$ bands. 
Furthermore, CB edge at the $\Gamma$ point which shows movment towards the E$_f$ for doping values x = 0.0 to 0.125 
shifts slightly in opposite direction (away from the E$_f$) for x = 0.25 and 0.5. 
This could be related to readjustment of the fnn Ge-p$_{z}$ orbital with respect to the fnn Ge-p$_{x/y}$
orbitals.

\section{Conclusion} 
We have presented a detail study of the  electronic structure evolution of Ge$_{1-x}$Pb$_x$Se, where x = 0.0, 0.125, 
0.25, 0.375, and 0.5  using first principles method. The Se-p and Ge-p states are strongly hybridized in
GeSe and a gap is observed in the DOS at E$_f$. This gap reduces systematically with simultaneous
enhancement of the states in the near E$_f$ region as a function of Pb doping.
The doping mainly affects the coordination environment around the doped atom site as clear
from the  relaxed structures of the doped compounds. This result is consistent with the DOS results
where we observe that the  near E$_f$ modified states are mainly composed of the fnn Se-p and Ge-p states. 
 Similarly, in the band structure plots we found shifting of the VB and CB edges which gives rise to noticeable decrease
in the indirect band in GeSe with Pb doping. This reduction of indirect band gap matches
qualitatively with the experimental results. From Wannier functions it is revealed that
the bond center of Pb-Se bond lies close to Se atom which gives rise to slight positive charge
at the doped atom site. The electrostatic effect originating from these doped atoms tries to
increase the BE of the fnn Ge-p$_{x/y}$ orbitals. Whereas, bonding electron cloud
centered at the fnn Se site screens the positive charge of the doped atom and reduces
the energy of the fnn Se-p$_{x/y}$ orbitals.   
Therefore,  this electrostatic influence of the doped Pb atom could be the reason for
shifting of the VB and CB edges and resulting indirect band gap 
reduction in Ge$_{1-x}$Pb$_x$Se  system.
Our comprehensive study is helpful for further experimental and theoretical investigations of group IV-VI compounds
in the context of controlling  their properties with doping.


\begin{thebibliography}{99}
\bibitem{not} Pseudopotentials utilized in the present calculations were obtained from the quantum espresso pseudopotential library.
\end{thebibliography}
\end{document}